\documentclass[12pt]{iopart}
\usepackage{graphicx}

\begin{document}

\paper{The quantum double well anharmonic oscillator in an external field}
\author{Erik Van der Straeten and Jan Naudts}

\address{Departement Fysica, Universiteit Antwerpen, \\Universiteitsplein 1, 2610 Antwerpen, Belgium}
\eads{\mailto{Erik.VanderStraeten@ua.ac.be}, \mailto{Jan.Naudts@ua.ac.be}}

\begin{abstract}
The aim of this paper is twofold. First of all, we study the behaviour of the lowest eigenvalues of the quantum anharmonic oscillator under influence of an external field. We try to understand this behaviour using perturbation theory and compare the results with numerical calculations. This brings us to the second aim of improving the method used to carry out the numerical calculations.
\end{abstract}

\section{Introduction}
The quantum anharmonic oscillator is an important model in physics. Unfortunately it can't be solved in closed form. In the past, intensive efforts were carried out to develop reliable methods to solve the model numerically. Many papers study only the double well potential \cite{referee1,referee2,referee3}. Other papers deal with all sorts of quantum anharmonic oscillators (see, e.g. \cite{referee4} and references therein). The main point of these papers is to develop numerical methods to solve the eigenvalue problem of anharmonic oscillators as accurately as possible. Selecting the best method out of all these different proposals is a hard task. 

The problem of the quantum anharmonic oscillator in an external field has not yet been studied in detail, although it has interesting applications. In \cite{referee5} one studies numerically chains of coupled quantum anharmonic oscillators. Such chains are used in condensed matter as one dimensional models of crystals and are a first interesting step towards quantum field theory. In this context, coupled anharmonic oscillators are used to study displacive and order-disorder phase transitions (see e.g. \cite{referee6}). The anharmonic oscillator in an external field appears in the mean field treatment of a chain of coupled anharmonic oscillators when the interactions between the oscillators are replaced by an effective external field.

In Section 2, we will follow the lines of \cite{referee7} to calculate the eigenvalues of the anharmonic oscillator with a general polynomial potential.  In Section 3, we apply the general results obtained in Section 2 to the anharmonic oscillator in an external field. In Section 4 and 5 we study the lowest eigenvalues of respectively a shallow and a deep well. The final section gives a short discussion of the results.

\section{Method}
We use a method which is based on two papers \cite{referee2,referee7}. In \cite{referee7}, Jafarpour and Afshar calculate the energy eigenvalues of the anharmonic oscillator with a general polynomial potential. To obtain these eigenvalues they define a 'normalised squeezed vacuum state' and use it to build up an orthonormal basis. With a variational approach, this basis is optimised. Then it is used to calculate the matrix representation of the Hamiltonian. Finally this matrix is diagonalised numerically to obtain the lowest energy eigenvalues. 

Following the lines of \cite{referee7} we calculate the eigenvalues of the general Hamiltonian
\begin{equation} \label{hamilgen} 
H=\frac{1}{2m}P^2+\sum_{i=0}^I\lambda_iQ^i.
\end{equation}
However, we adapt the method  of \cite{referee7} at two important places.
First, we use the well known basis of the harmonic oscillator in stead of the transformed basis introduced in \cite{referee7}.
This makes the method more transparent. Second, we use a more general method for the optimisation of the choice of basis.

\begin {description}

\item {1)}
Our modification of the method starts from the observation that the 'normalised squeezed vacuum state'  of \cite{referee7}
is just the ground state of a harmonic oscillator with suitably chosen parameters.
Indeed, the transformation from one harmonic oscillator to another one, with different mass $m$ and frequency $\omega_0$,
is a special case of a Bogoliubov transformation. The generic case of the Bogoliubov transformation is not used in \cite{referee7}.
The basis functions used by the authors are the well known eigenfunctions of the harmonic oscillator, be it with
optimised value of the parameters $m$ and $\omega_0$. The method of \cite{referee7} can therefore be simplified by omitting the
use of Bogoliubov transformations. 

\item {2)}
In the literature, other methods \cite{referee1,referee2} are found than that of \cite{referee7}, to optimise the choice of orthonormal basis. We will discuss some of these methods and use the one best suited for the problem at hand.

\end {description}

A more detailed description of the method now follows.

The operators $a$ and $a^\dagger$ are the annihilation and creation operators of the ordinary harmonic oscillator with mass $m$ and frequency $\omega_0$. With these operators and the notation $m\omega_0r^2=\hbar$ Hamiltonian (\ref{hamilgen}) becomes
\begin{equation}
H=-\frac{\hbar^2}{4mr^2}\left(a-a^\dagger\right)^2+\sum_{i=0}^I\lambda_i\left(\frac{r}{\sqrt 2}\right)^i\left(a+a^\dagger\right)^i.
\end{equation}
An ordered expression for $\left(a+a^\dagger\right)^i$ is derived in \cite{referee7}. Using this expression the general Hamiltonian can be written as
\begin{equation}
\fl H=-\frac{\hbar^2}{4mr^2}\left(a-a^\dagger\right)^2+\sum_{i=0}^I\lambda_i\left(\frac{r}{\sqrt 2}\right)^i\sum_{k=0}^{\lfloor i/2\rfloor}\frac{i!}{2^kk!}\sum_{j=0}^{i-2k}\frac{(a^\dagger)^{i-2k-j}a^j}{j!(i-2k-j)!}.
\label {cranh}
\end{equation}
$\lfloor x\rfloor$ means rounding the value of $x$ to the lower integer.

The method proceeds in three steps. First, the expectation value of the Hamiltonian (\ref {cranh}) is calculated in some eigenstate of the harmonic oscillator (the choice of eigenstate will be discussed later on). Second, the expectation value is minimised by varying the parameter $\omega_0m$, or equivalently $r$, of the harmonic oscillator. Third, the eigenfunctions of the harmonic oscillator (now with fixed parameter) are used to obtain the matrix representation of the Hamiltonian $H$.

The expectation value in an arbitrary eigenstate, labeled with parameter $t$, is given by
\begin{equation} \label{expvalue}
\fl \left<\psi_t\right|H\left|\psi_t\right>=\frac{\hbar^2}{4mr^2}(2t+1)+\sum_{i=0,2}^I\sum_{k=0}^{\lfloor i/2\rfloor}\frac{\lambda_ir^ii!t!}{2^{i/2+k}k!((i/2-k)!)^2(t-i/2+k)!}.
\end{equation}
After minimising this expectation value ($\partial \left<\psi_t\right|H\left|\psi_t\right>/\partial r^2$=0), the following equation is obtained
\begin{equation} \label{vwdtwee}
\fl 0=-\frac{\hbar^2}{4mr^4_0}(2t+1)+\sum_{i=2,4}^I\sum_{k=0}^{\lfloor i/2\rfloor}\frac{i\lambda_ir_0^{i-2}i!t!}{2^{i/2+k+1}k!((i/2-k)!)^2(t-i/2+k)!}.
\end{equation}
This equation gives one condition to calculate the value of two parameters ($r_0$ and $t$). In practice we will chose a value for the parameter $t$, and solve the equation to obtain the value of $r_0$, the optimal value of $r$. In general this equation has to be solved numerically. In the following section, we return to the problem of choosing the parameter $t$. 

The eigenfunctions of the harmonic oscillator with $r=r_0$ are now used as a basis. The matrix elements of Hamiltonian (\ref {cranh}) become
\begin{eqnarray} \label{matrixH}
\fl \left<\psi_s\right|H\left|\psi_t\right>=\frac{\hbar^2}{4mr^2_0}\left[(2t+1)\delta_{s,t}-\sqrt{t(t-1)}\delta_{s,t-2}-\sqrt{(t+1)(t+2)}\delta_{s,t+2}\right]
\nonumber\\
\lo +\sum_{i=0}^I\sum_{k=0}^{\lfloor i/2\rfloor}\sum_{j=0}^{i-2k}\lambda_i\frac{r_0^i}{2^{i/2}}\frac{i!}{2^kk!}\frac{\sqrt{t!(t+i-2j-2k)!}}{j!(t-j)!(i-2k-j)!}\delta_{s,t+i-2j-2k}.
\end{eqnarray}
Note that the last part equals zero, if $j>t$ and $i-2k-j>s$. 
The matrix is infinite dimensional. Truncating the basis to $N$ basis functions results in an $N$ dimensional matrix. After diagonalising this $N$ dimensional matrix, the low-lying eigenvalues of the Hamiltonian are found.

\section{Anharmonic oscillator in an external field}
From now on we restrict ourselves to the quantum anharmonic oscillator in an external field. The Hamiltonian is given by
\begin{eqnarray} \label{hamil1}
H=\frac{1}{2m}P^2+V\left(Q\right)-pQ=\frac{1}{2m}P^2+\frac{\alpha}{2}Q^2+\frac{\beta}{4}Q^4-pQ,
\end{eqnarray}
with $\alpha<0$ and $\beta>0$. The potential $V\left(Q\right)$ is a double well potential. The extra term $-pQ$ favours one of the two wells. 

Rewriting this Hamiltonian to the form of (\ref{hamilgen}) gives $I=4$, $\lambda_0=\lambda_3=0$, $\lambda_1=-p$, $\lambda_2=\alpha/2$ and $\lambda_4=\beta/4$. Then equation (\ref{vwdtwee}) has to be solved, to obtain the value of $r_0$. In this case, the equation reduces to
\begin{equation} \label{recon}
(2t+1)\left(2\lambda_2r_0^4-\frac{\hbar^2}{m}\right)+6\lambda_4r_0^6(2t^2+2t+1)=0.
\end{equation}
This is a cubic equation in $r_0^2$ which has only one real and positive solution (remember $\alpha<0$).

For the double well potential different choices of the parameter $t$ are proposed in the literature. None of these choices has a physical interpretation. The only reason to prefer one choice above another is a faster convergence of the numerical results as a function of $N$. 

The condition derived in \cite{referee7} corresponds with the choice $t=0$. This is not the optimal choice. The authors are aware of this and do not always use the calculated value of $r_0$ to obtain the numerical data. In \cite{referee1} Balsa et al.~minimise the expectation value (\ref{expvalue}) with respect to  the parameters $r$ and $t$. This way, the authors get two conditions for two parameters and no arbitrary choice is necessary. Problem is, as outlined in \cite{referee2} by Bishop et al., that the calculated values of the parameters are far from the optimal values. 

Bishop et al.~also noticed that the optimal choice of $r_0$ depends of $N$. This motivates the choice $t=N/2$. This is the choice we will use to obtain our numerical results. As mentioned earlier, there is no physical justification for this choice. The only motivation is that we get a fast convergence of the numerical results as a function of $N$ for every example we studied.

We are interested in the behaviour of the eigenvalues under influence of an external field. So we have to repeat the diagonalisation of the Hamiltonian several times for different values of the parameter $p$. Note that we don't have to repeat the calculation of $r_0$ because equation (\ref{recon}) doesn't depend on $p$.

\section{Shallow well}
\begin{figure}[t]
\parbox{7cm}{\includegraphics[width=7cm]{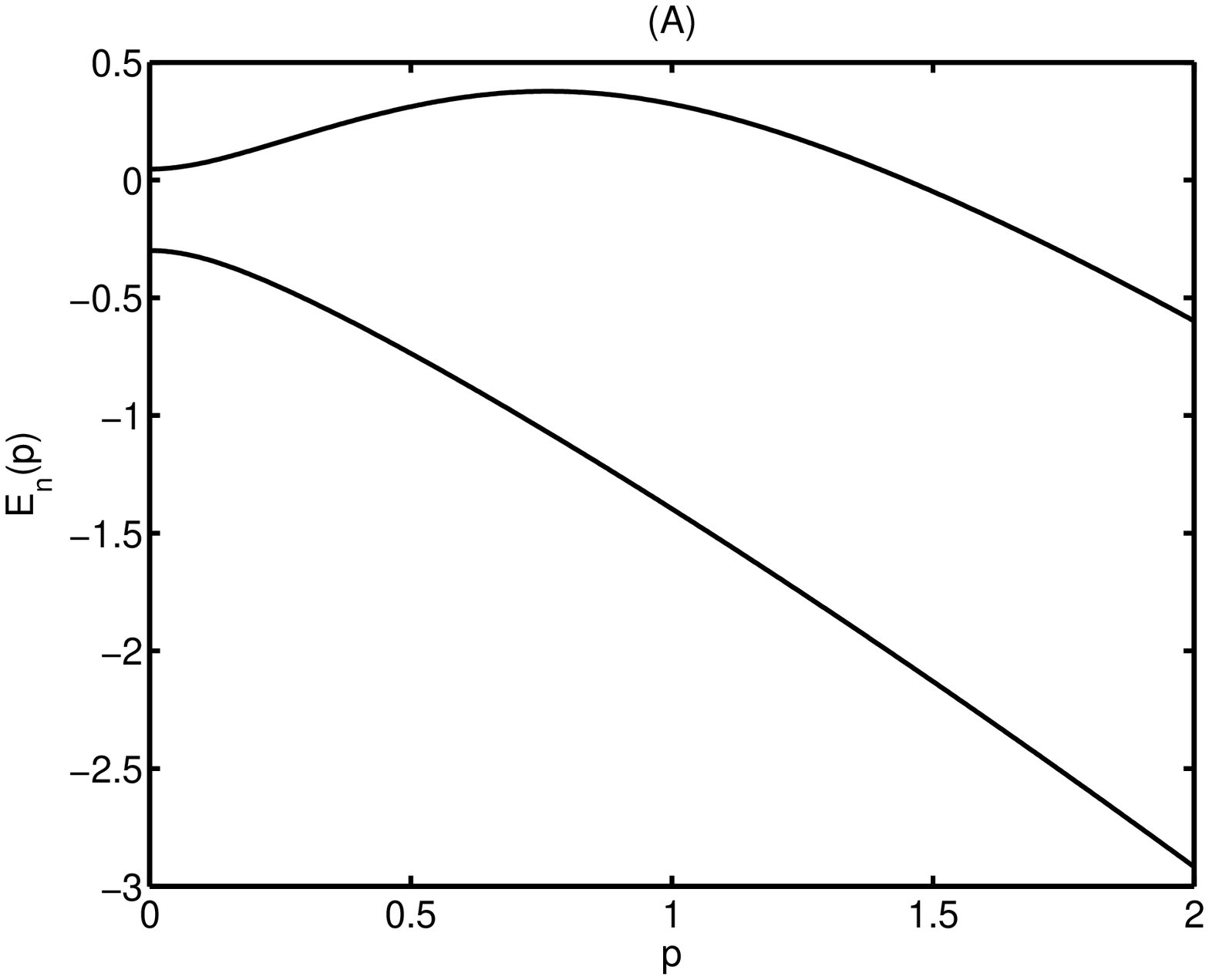}}
\hfill
\parbox{7cm}{\includegraphics[width=7cm]{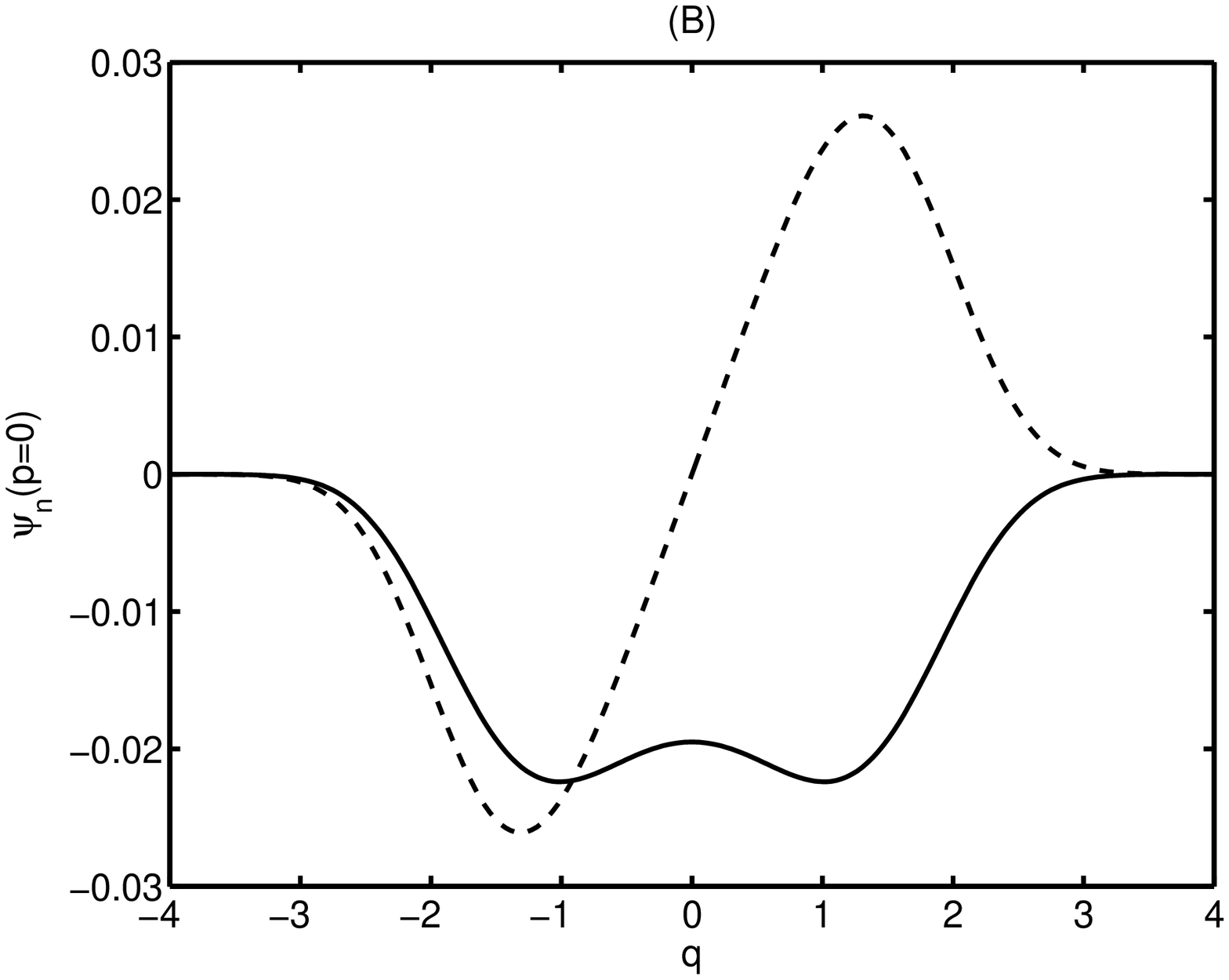}}
\caption{\label{figuur1}(A) First two eigenvalues of the anharmonic oscillator in an external field as a function of the parameter $p$, with $N=30$. (B) Position representation of the wavefunctions of the ground state (\full) and the first excited state (\dashed) of the anharmonic oscillator without external field ($p=0$), with $N=30$.}
\end{figure}
For the first example we choose $\alpha=-2$ and $\beta=1$.

Figure (\ref{figuur1}A) shows the eigenvalue of the ground state and the first excited state as a function of the parameter $p$. If $p=0$, the potential is the symmetric double well potential. If an extra term $-pQ$ is added, the energy decreases, because the system can lower it's energy by partly choosing one of the wells. However, the energy of the first excited state increases for small values of $p$. To understand this, we look for the response of the wavefunctions to the parameter $p$.

\begin{figure}[t]
\parbox{7cm}{\includegraphics[width=7cm]{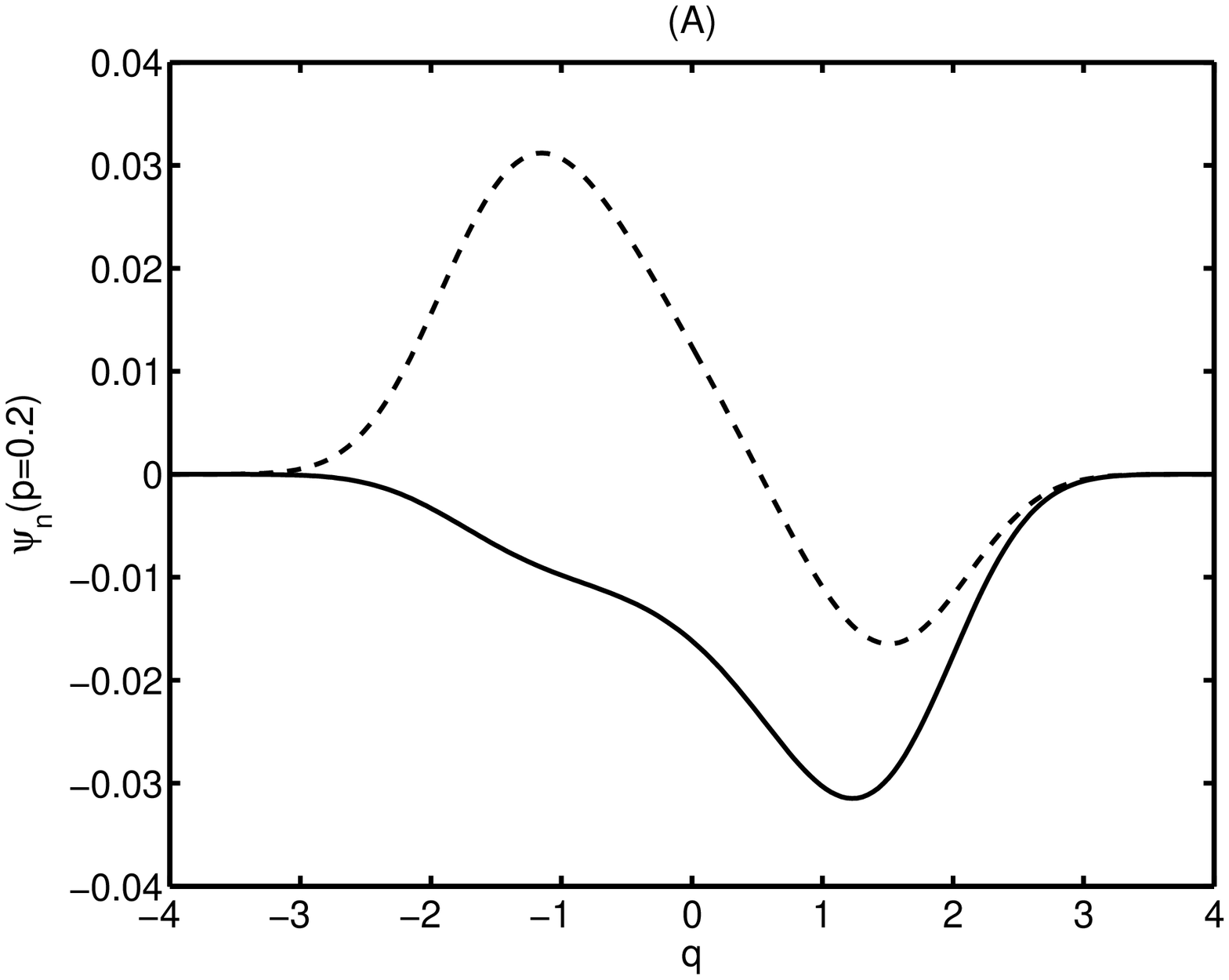}}
\hfill
\parbox{7cm}{\includegraphics[width=7cm]{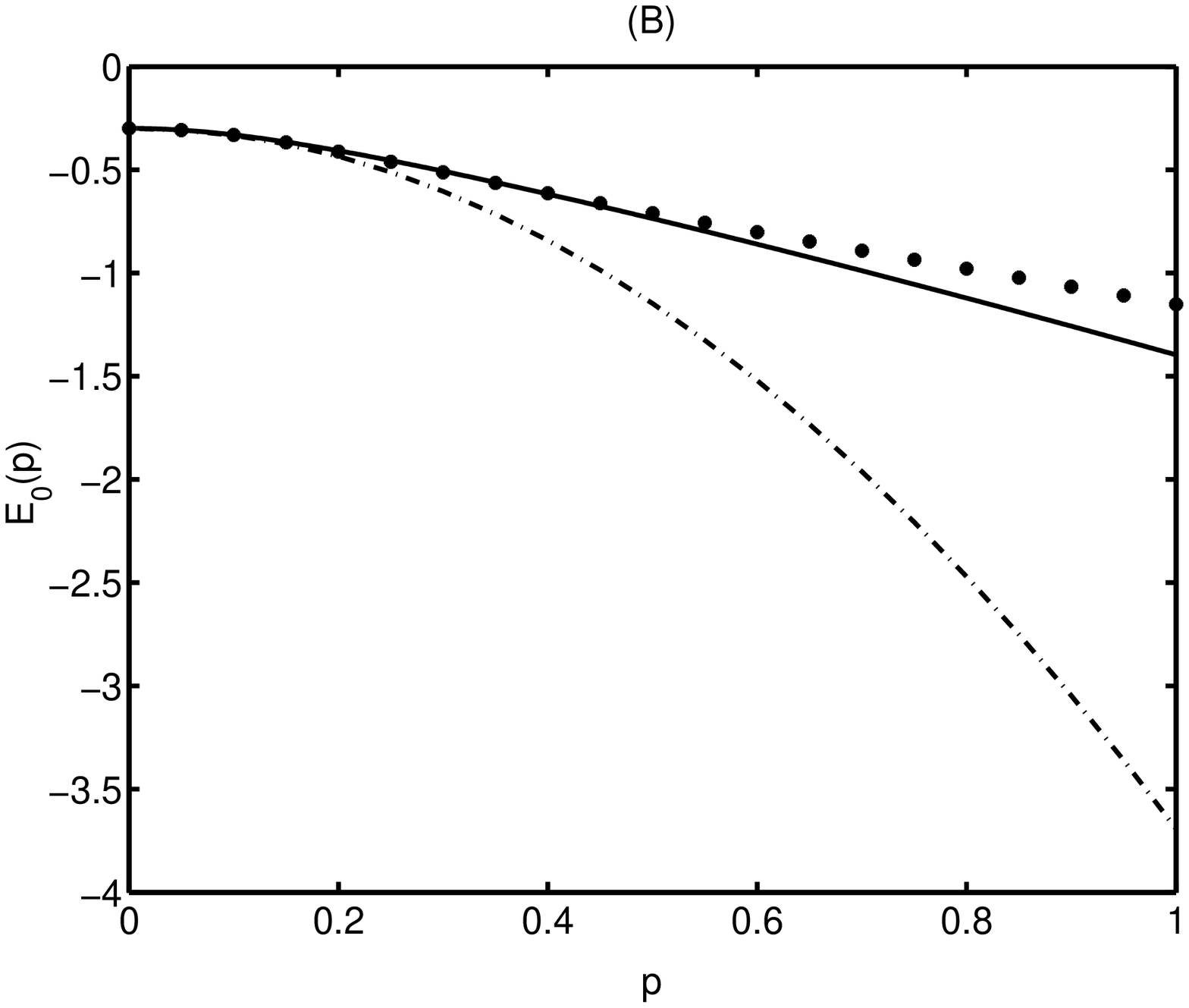}}
\caption{\label{figuur2}(A) Position representation of the wavefunctions of the ground state (\full) and the first excited state (\dashed) of the anharmonic oscillator in an external field ($p=0.2$), with $N=30$. (B) Comparison of the result of equation (\ref{approxv}) (\dashed), equation (\ref{fiteqn}) (\fullcircle)  and the original numerical data (\full) for the ground state. We used following values $c_1=-3.3921$, $a=-0.8531$, $\omega=c_1/a=3.9762$ and $N=40$.}
\end{figure}

Figure (\ref{figuur1}B), shows the wavefunction of the ground state and the first excited state in the position representation, with $p=0$. Clearly, the moduli of the wavefunctions of the ground state and of the first excited state are both symmetric. If $p\neq 0$ this symmetry is destroyed.  Figure (\ref{figuur2}A), shows the wavefunctions of the ground state and the first excited state in the position representation, with $p=0.2$. We see that the wavefunction of the ground state has a large peak for positive values of $q$. Because of the orthogonality of the wavefunctions, the wavefunction of the first excited state must have it's largest contribution for negative values of $q$. As a consequence, the energy of the ground state will decrease and the energy of the first excited state will increase.

Let us now focus on the response of the ground state eigenvalue to an external field. We consider the external field as a perturbation of the anharmonic oscillator. According to non-degenerate perturbation theory to second order, the energy of the ground state is given by
\begin{equation} \label{alguit}
E_0(p)\approx E_0(0)-p\left<\psi_0(0)\right|Q\left|\psi_0(0)\right>+p^2\sum_{m(\neq 0)}\frac{\left|\left<\psi_0(0)\right|Q\left|\psi_m(0)\right>\right|^2}{E_0(0)-E_m(0)}.
\end{equation}
The modulus of $\left|\psi_0(0)\right>$ is symmetric in $q$. As a consequence, the linear term in expression (\ref{alguit}) vanishes and the leading order terms of expression (\ref{alguit}) are
\begin{equation} \label{approxv}
E_0(p)\approx E_0(0)+p^2\frac{\left|\left<\psi_0(0)\right|Q\left|\psi_1(0)\right>\right|^2}{E_0(0)-E_1(0)}=E_0(0)+c_1p^2,
\end{equation}
because $E_0(0)-E_1(0)<<E_0(0)-E_i(0)$ for $i>1$.
The results of a numerical calculation of $E_0(0),\ E_1(0)$ and $c_1$ for different values of $N$ are given in table (\ref{tabel1}A). As is clearly seen in figure (\ref{figuur2}B), equation (\ref{approxv}) is only valid for very small values of the parameter $p$.

For intermediate values of the parameter $p$, the tunnel splitting between the ground state and the first excited state can be ignored. As a consequence, the ground state becomes twofold degenerate. Degenerate perturbation theory gives
\begin{equation}
E_0(p)\approx E_0(0)-p\left|\left<\psi_0(0)\right|Q\left|\psi_1(0)\right>\right|=E_0(0)-p\left|Q_{01}\right|.
\end{equation}
A numerical calculation of this matrix element gives following approximate value $\left|Q_{01}\right|\approx0.853104$.

We therefore expect that the ground state can be fitted for small and intermediate values of the parameter $p$ by following equation
\begin{equation} \label{fiteqn}
E_0(p)\approx E_0(0)+ap\tanh(\omega p).
\end{equation}
\fulltable{\label{tabel1} The values of $r_0^2$ are calculated with equation (\ref{recon}) and the choice $t=N/2$. In this table, the values of $r_0^2$ are rounded but for the numerical calculations we used the exact solution of this equation. (A) The results of a numerical calculation of $E_0(0),\ E_1(0)$ and $c_1$, with $p=0$ and for different dimensions of the Hamiltonian matrix. (B) Fitting parameter for different dimensions of the Hamiltonian matrix.}
\br
 &&(A)&&&(B)
 \\
$N$  & $r_0^2$ & $E_0$ & $E_1$ & $c_1$ & $a$
\\
\mr
 10 & 0.59 & \-0.299479413549 & 0.046558837188 & \-3.390313017166 &  \-0.930574586052
\\
 20 & 0.45 & \-0.299521364979 & 0.046371082733 & \-3.392128162181 &  \-0.930891427590
\\ 
 30 & 0.38 & \-0.299521367416 & 0.046371082228 & \-3.392128193573 &  \-0.930891443755
\\
 40 & 0.34 & \-0.299521367416 & 0.046371082228 & \-3.392128193573 &  \-0.930891444167
\\
\br
\endfulltable
For the limit $p\rightarrow 0$ the desired quadratic behaviour $E_0(p)\approx E_0(0)+a\omega p^2$ is obtained. For large $p$, $\tanh(\omega p)$ goes to one exponentially fast. There remains a linear equation $E_0(p)\approx E_0(0)+ap$. 

We fix the value of $\omega$ by the condition $a\omega=c_1$. The only remaining fitting-parameter is then $a$. The values of $a$ for different dimensions of the Hamiltonian matrix are shown in table (\ref{tabel1}B). The value of $a\approx-0.931$ is approximately the same as the value of $-\left|Q_{01}\right|$ mentioned above. This means that it is a good approximation to neglect the tunnel splitting for intermediate values of $p$.

Figure (\ref{figuur2}B) shows the result of equation (\ref{approxv}) and equation (\ref{fiteqn}) together with the original numerical data (with $N=40$). The figure demonstrates that the non-degenerate perturbation theory holds for very small values of the parameter $p$. Equation (\ref{fiteqn}) holds for small and intermediate values of $p$. However, it is clear that for large values of the parameter $p$ higher order terms become important. We conclude that equation (\ref{fiteqn}) in combination with the results of the perturbation theory ($a=-\left|Q_{01}\right|$ and $a\omega=c_1$) is a good approximation for small and intermediate values of $p$. No fitting-parameters are necessary.

For very large values of the parameter $p$, the depth of the well is asymptotically given by $-0.4725p^{4/3}/\lambda_4^{1/3}$. We therefore expect for very large $p$-values the following behaviour of the ground state energy
\begin{equation}
E_0(p)\approx A+Bp^{4/3},
\end{equation}
with $A$ and $B$ constants. This relation is in agreement with numerical calculations.

\section{Deep well}
\begin{figure}[t]
\parbox{7cm}{\includegraphics[width=7cm]{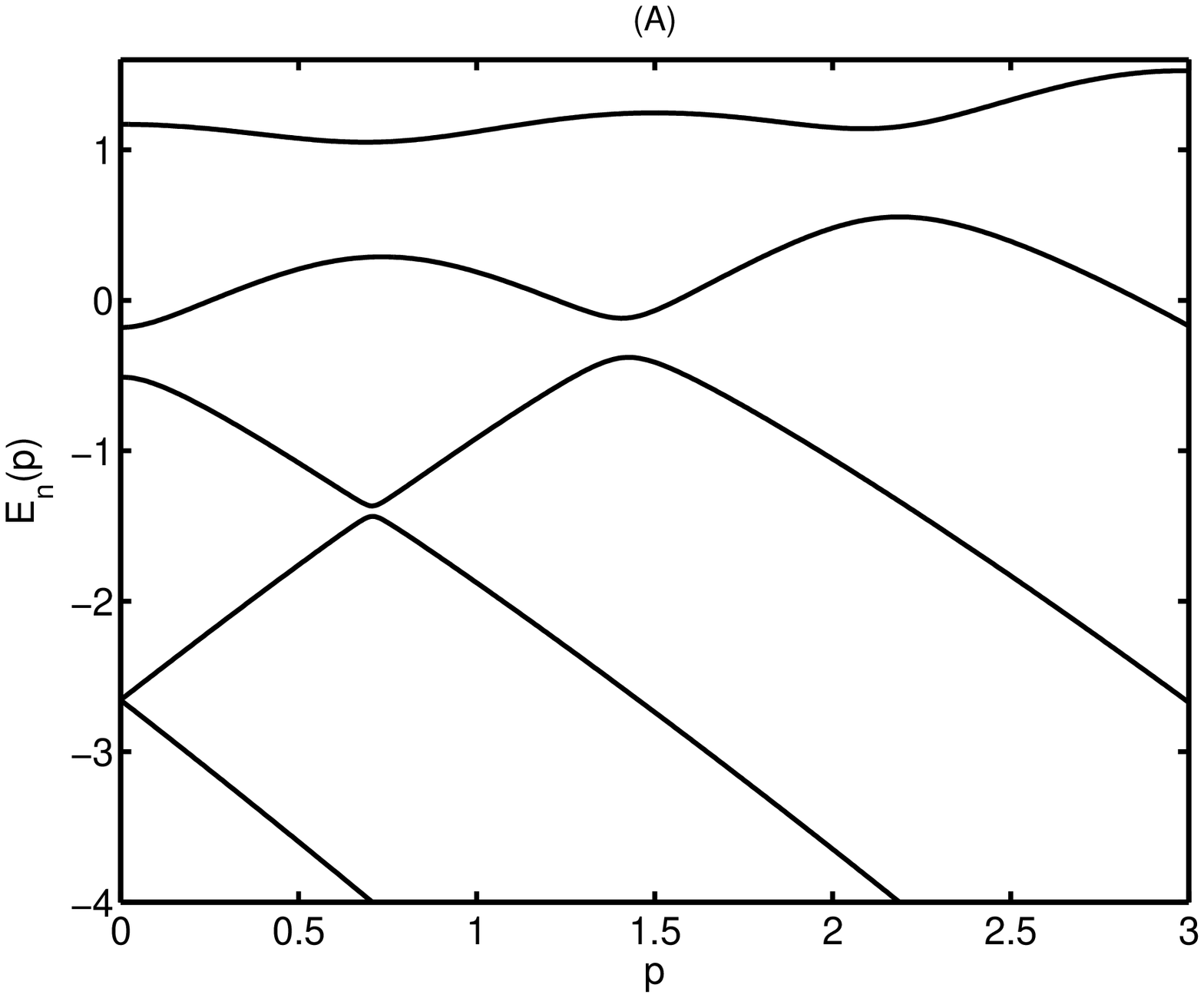}}
\hfill
\parbox{7cm}{\includegraphics[width=7cm]{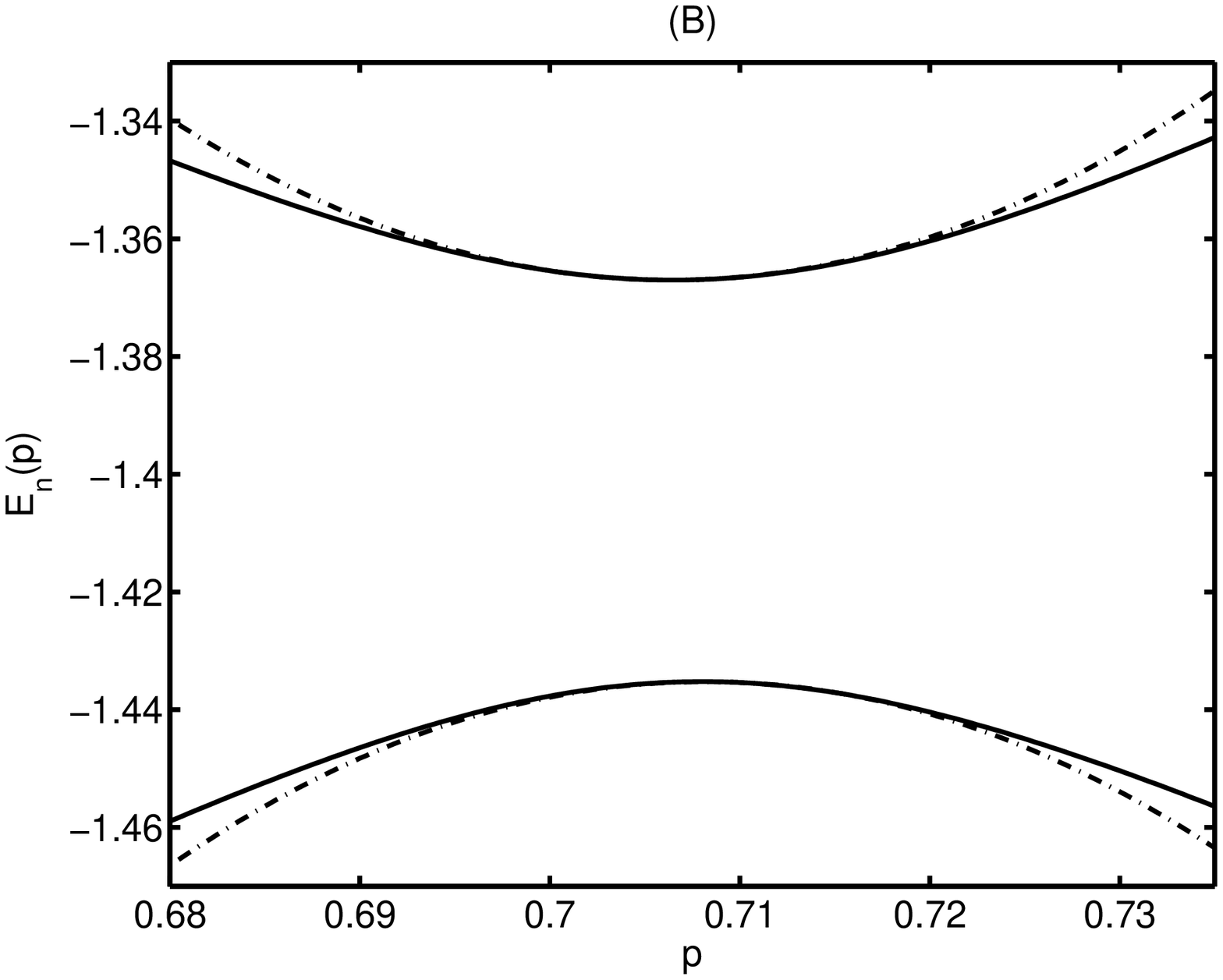}}
\caption{\label{figuur3}(A) First five eigenvalues of the anharmonic oscillator in an external field as a function of the parameter $p$. (B) Comparison of the result of equation (\ref{e1fit}) and equation (\ref{e2fit}) (\dashed)  and the original numerical data (\full) in the surroundings of $p_1$. We used following values $Q_{11}=-0.05912$, $c_2=-39,30905$ and $Q_{22}=-0.05902$.}
\end{figure}
For this second example we choose $\alpha=-4$ and $\beta=1$. All numerical calculations for this example are carried out with $N=50$.

Figure (\ref{figuur3}A) shows the first five eigenvalues as a function of the parameter $p$. This figure demonstrates the repulsion of nearly degenerate energy levels. We can understand this from perturbation theory. Let us focus on the repulsion of the first and second excited state. The value of $p$ where the energy difference between these two states is minimal, is denoted $p_1$ and approximately equals $0.70724$. For $E_1$, the leading order terms of perturbation theory around $p_1$ are 
\begin{eqnarray} \label{e1fit}
\fl E_1(p_1+\Delta p)\approx E_1(p_1)+\Delta p\left<\psi_1(p_1)\right|Q\left|\psi_1(p_1)\right>+(\Delta p)^2\frac{\left|\left<\psi_1(p_1)\right|Q\left|\psi_2(p_1)\right>\right|^2}{E_1(p_1)-E_2(p_1)}
\nonumber\\
\lo =E_1(p_1)+Q_{11}\Delta p+c_2(\Delta p)^2.
\end{eqnarray}
The linear term does not vanish because the modulus of $\left<\psi_n(p)\right|$ is not symmetric for $p\neq0$. A similar expression can be derived for $E_2$
\begin{equation} \label{e2fit}
E_2(p_1+\Delta p)\approx E_2(p_1)+Q_{22}\Delta p-c_2(\Delta p)^2.
\end{equation}
The values of $E_1(p)$ and  $E_2(p)$ are almost equal in $p_1$. As a consequence the linear terms of equation (\ref{e1fit}) and equation (\ref{e2fit}) are also almost equal. Because of the opposite signs of the quadratic terms in these equations, the perturbation will lower one energy level and raise the other.

A numerical calculation gives the following approximate values $Q_{11}\approx -0.05912$, $c_2\approx -39.30905$ and $Q_{22}\approx -0.05902$. Figure (\ref{figuur3}B) shows the result of equation (\ref{e1fit}) and equation (\ref{e2fit}) together with the original numerical data in the surroundings of $p_1$. It's clear that the perturbation series give good approximations for the energy levels in the surroundings of $p_1$.

\section{Discussion}
We study the behaviour of the lowest eigenvalues of the quantum anharmonic oscillator under influence of an external field. The method used to obtain our numerical data is based on two papers \cite{referee2,referee7}. We follow the lines of \cite{referee7} to calculate the eigenvalues of the general Hamiltonian (\ref{hamilgen}). We use the eigenfunctions of the harmonic oscillator as a basis. We optimise this choice of basis by varying a parameter of the harmonic oscillator, using a criterion inspired by \cite{referee2}. Finally this matrix is diagonalised numerically. 

We apply the method to study the anharmonic oscillator in an external field. In \cite{referee7}, Jafarpour and Afshar considered the anharmonic oscillator as a test example. We were able to reproduce their results but with a quicker convergence.

We are especially interested in the response of the energy of the ground state to an external field. The anharmonic oscillator in an external field is also used in \cite{referee3} as a test example for still an other technique to calculate the eigenvalues of anharmonic oscillators. The authors looked also to the response of the energy levels to an external field, but only for intermediate values of the external field. In our opinion, there are three different regimes. In the first regime (small external field), we use non-degenerate perturbation theory to explain the quadratic response of the energy of the ground state. In the second regime (intermediate external field), we use degenerate perturbation theory to explain the linear response of the energy of the ground state (just like in \cite{referee3}). In the third regime (large external field), higher order terms become important. We propose formula (\ref{fiteqn}), which is able to produce the desired behaviour in the combined regime of small and intermediate external fields. This formula is supported by numerical calculations, as is clearly visible in figure (\ref{figuur2}B). 

In \cite{referee3}, the authors also noticed the repulsion of the nearly degenerate energy levels. Again they only looked at intermediate external fields. They explained with degenerate perturbation theory why the energy levels become nearly equal. The repulsion of the nearly degenerate energy levels is explained by non-degenerate perturbation theory. We show that the repulsion is caused by the opposite sign of the quadratic term in the perturbation series. This is confirmed by numerical calculations and is clearly visible in figure (\ref{figuur3}B). 

In conclusion, we have shown that the present method can be used to calculate accurately the eigenvalues of the anharmonic oscillator in an external field. We also have shown that non-degenerate and degenerate perturbation theory are needed to explain the response of the energy of the ground state to an external field. 

In future work, we want to use these results to study the semi-classical behaviour of the quantum anharmonic oscillator.

\end{document}